\documentclass[twocolumn,eqsecnum,showpacs,preprintnumbers,amsmath,amssymb]{revtex4}
\def\shiftdown#1{#1\llap{\lower.04ex\hbox{#1}}}

\begin{document}
\title{Hyperquarks and generation number}
\renewcommand{\thefootnote}{\fnsymbol{footnote}}
\author{Alfons J. Buchmann and Michael L. Schmid}
\affiliation{Institut f\"ur Theoretische Physik, Universit\"at T\"ubingen,
         Auf der Morgenstelle 14, D-72076 T\"ubingen, Germany}
\email{alfons.buchmann@uni-tuebingen.de}   

\email{micha.l.schmid@gmx.net} 

\begin{abstract}
In a model in which quarks and leptons are built up from two
spin-$\frac{1}{2}$ preons as fundamental entities,
a new class of fermionic bound states (hyperquarks) arises.
It turns out that these hyperquarks are 
necessary to fulfill the 't Hooft anomaly constraint,
which then links the number of fermionic generations 
to the number of colors and hypercolors.
\end{abstract}

\smallskip

\maketitle

\section{Introduction}
\label{sec:1}

The fact that leptons and quarks form families of identical structure and 
the large number of adhoc parameters of the standard model have led to the 
conjecture that these particles are composites of more fundamental entities. 
Among the various approaches to quark and lepton substructure, 
the Harari-Shupe model, which has only two elementary spin-$\frac{1}{2}$ 
building blocks stands out because of its simplicity.  
For the construction of leptons and quarks, one of these
($T$-preon) carries the fundamental electric charge of 
$\left(\frac{e}{3}\right)$, the other one ($V$-preon) is 
neutral~\cite{har79,shu79}. 

The preon model should give answers to
some fundamental questions, which remain unanswered within the standard model,
for example:
\begin{itemize}
\item[(i)] How can the exact equality of the electric charge of proton 
and positron be explained? 
\item[(ii)] Is there a connection between fractional charges and 
colored fermions? 
\item[(iii)] Why are there exactly three identical fermionic families 
(generations)? 
\end{itemize}
While answers to the first two questions 
have been suggested, there has not been much progress in answering the third one.
The purpose of this paper is to propose an answer to the question why there are 
three generations of quarks and leptons.  

With respect to preon dynamics not much is known except that ordinary 
quantum chromodynamic (QCD) forces alone are insufficient to bind preons 
into both colored quarks and colorless leptons. For preon binding 
a new superstrong interaction (hypercolor force) at the TeV range~\cite{har80} 
was suggested. Thus, preons carry both hypercolor and color charges. 
Quarks and leptons are assumed to be hypercolor singlets. The corresponding 
gauge group has an analogous structure as the SU(3) color group of QCD and 
is called quantum hyperchromodynamics~\cite{har80,har82a}. The model involves 
three unbroken local gauge symmetries: hypercolor, color, and electromagnetism, 
and the total gauge group is SU(3)$_{H}$ $\times$ SU(3)$_{C}$ $\times$ U(1)$_{Q}$.  

The small radius of the quarks and leptons, $r < 10^{-19}~\mbox{m} $, corresponds 
to a preon kinetic energy of $T > \mbox{1 TeV} $.
Most of this kinetic energy must be compensated by a nearly equal potential 
energy of the superstrong forces, in order to explain the near masslessness of 
leptons and quarks compared to the effective preon mass. The chiral symmetry at 
the preon level must not be spontaneously broken at the energy scale of the bound 
states~\cite{gth79}. Otherwise, the bound state masses would be of the same order of 
magnitude as the effective preon mass $\left( > \mbox{1 TeV}\right)$.
For example, in QCD, where chiral symmetry is spontaneously broken, bound state 
masses are of the same order of magnitude as the constituent quark masses.

In order to prevent chiral symmetry to be spontaneously broken, a chiral 
anomaly cancellation condition, the ' t Hooft anomaly constraint, must be 
satisfied~\cite{gth79}. The Harari-Shupe model with only hypercolor singlets 
does not satisfy this condition. In this paper we introduce a new class of
bound states (hyperquarks) which reconciles the Harari-Shupe model with the 
't Hooft anomaly constraint. Furthermore, we suggest a simple argument which 
links the number of fermion families to the number of colors and hypercolors.

\section{Preon quantum numbers}
\label{sec:2}

In the Harari-Shupe model all quarks and leptons can be built from just 
two spin-$\frac{1}{2}$ fermions (preons). According to Harari and 
Seiberg~\cite{har80} the two types of preons belong to the following 
representations of the underlying exact gauge group SU(3)$_{H}$ 
$\times$ SU(3)$_{C}$ $\times$ U(1)$_{Q}$: T: (3,3)$_{\frac{1}{3}}$ and V: (3,$\bar{3}$)$_{0}$, 
where the first (second) argument is the dimension of the representation in
hypercolor (color) space and the subscript denotes the electric charge. 
Although there are two degenerate types of preons (T and V) there is no global SU(2) 
isospin symmetry on the preon level because the charged and neutral preon belong to different 
representations in color space.  To be consistent with the parity assignment 
for the standard model 
fermions, the intrinsic parity of the $T$ and $V$ preons must be different. 
The preon quantum numbers are summarized in Table~\ref{tab:1}. 

\begin{table}
\begin{center}
\begin{tabular}{l c c c c c c}
\hline
preon \ \ \ \  & H \ \ \  & C  \ \ \ & Q \ \ \  & ${\cal P}$ \ \ \ & $ \Upsilon $ \ \ \  & $\Pi$ \\
\hline
& & & & & &  \\
$T$ \ \ & $3$ \ \ \ & $3$ \ \ \  & $ +\frac{1}{3} \ \ \ $ & $+\frac{1}{3}$ \ \ \ & $+\frac{1}{3}$ \ \ \  & -1 \\
& & & & & &  \\
\hline
& & & & & &  \\
$V$ \ \ & $3$ \ \ \ & $\bar{3}$ \ \ \ & 0 \ \ \ & $+\frac{1}{3}$ \ \ \ & $-\frac{1}{3}$ \ \ \ & +1  \\    
& & & & & &  \\
\hline
& & & & & & \\
$\bar{V}$ \ \ \ & $\bar{3}$ \ \ \ & $3$ \ \ \ & 0  \ \ \ & $-\frac{1}{3}$ \ \ \ & $+\frac{1}{3}$ \ \ \ & -1 \\
& & & & & & \\
\hline
& & & & & & \\
$\bar{T}$ \ \ \ & $\bar{3}$ \ \ \ & $\bar{3}$ \ \ \ & $ -\frac{1}{3} $ \ \ \ & $-\frac{1}{3}$ \ \ \ & 
$-\frac{1}{3}$  \ \ \ & +1 \\
& & & & & & \\
\hline
\end{tabular}
\caption{The color (C), hypercolor (H), electric charge (Q), preon number
${\cal P}$, $\Upsilon$ number, and intrinsic parity ($\Pi$)  of preons and antipreons 
(see also~\cite{har81}).} 
\label{tab:1}
\end{center}
\end{table}

Preons are characterized by new quantum numbers~\cite{har80}.
These are the preon number ${\cal P}$ and $\Upsilon$ number, 
which are combinations  of the number of T-preons n(T) and the number of V-preons n(V):
\begin{eqnarray}
\label{preonnumber}
{\cal P} &=& \frac{1}{3} \left(n(T) + n(V)\right) \\
\Upsilon &=& \frac{1}{3} \left(n(T) - n(V)\right). \nonumber 
\end{eqnarray}
The factor $\frac{1}{3}$ in Eq.(\ref{preonnumber}) is a convention. 
The $\Upsilon$ number is also related to the baryon number (B) and lepton 
number (L) of the standard model as $\Upsilon = (B - L)$.  
The antipreon numbers $n(\bar{T}$) and $n(\bar{V}$) are defined as 
$n(\bar{T}) = - n(T)$ and $n(\bar{V}) = - n(V)$.

There is a connection between the ${\cal P}$ and $\Upsilon$ numbers, 
and the electric charge $Q$ of the preons
\begin{equation}
\label{gellmannnishijima}
Q = \frac{1}{2} \left({\cal P} + \Upsilon\right),
\end{equation}
which can be readily verified from Table~\ref{tab:1}. 
This generalized Gell-Mann-Nishijima relation does not only hold for the 
preons but for all bound states (leptons, quarks, mesons, baryons)~\cite{har79}.

Furthermore, there is an interesting connection between these quantum numbers 
and the color (hypercolor) gauge symmetries~\cite{har83}. We generalize 
the definitions of ${\cal P}$ and $\Upsilon$ in Eq.(\ref{preonnumber}) 
for a theory with N$_{H}$ and N$_{C}$ (hyper)color charges of the preons
as follows: 
\begin{eqnarray}
{\cal P} &=& \frac{1}{N_{H}} \left(n(T) + n(V)\right) \nonumber \\ 
\Upsilon &=& \frac{1}{N_{C}} \left(n(T) - n(V)\right). 
\end{eqnarray}
After inserting these relations into Eq.(\ref{gellmannnishijima}) we obtain
for any multipreon state of charge $Q$ 
\begin{equation}
Q = \frac{1}{2} \left( {\cal P} + \Upsilon \right) = Q(T)n(T) + Q(V)n(V), 
\end{equation}
where
\begin{eqnarray} 
Q(T) &=& \frac{1}{2} \left( \frac{1}{N_{H}} + \frac{1}{N_{C}} \right) \nonumber \\ 
Q(V) &=& \frac{1}{2} \left( \frac{1}{N_{H}} - \frac{1}{N_{C}} \right).
\end{eqnarray}
Therefore, the equality of the number of colors and hypercolors leads
to electric neutrality of the V-preon 
\begin{equation}
Q(V) = 0.
\end{equation}
From the preceeding equations follows that electric charge is quantized because 
the ${\cal P}$ and  $\Upsilon$ numbers occur only as integer multiples of 1/3.

\section{Fermionic bound states}
\label{sec:3}

When constructing the three preon bound states, a new class of states
arises in addition to the usual leptons and quarks. 
We call these states hyperquarks~\cite{comment1}.  Hyperquarks have fractional 
electric charges similar to the quarks but carry hypercolor.
We make the following assumptions concerning possible bound states:
\begin{itemize}
\item[(i)] fermionic bound states consist of three preons, 
\item[(ii)] the bound states are singlets either in hypercolor (quarks), 
or color (hyperquarks), or both (leptons); 
only preons carry both hypercolor and color. 
\end{itemize}
We assume that these are the only allowed 
combinations.~$^{\footnotemark[4]}$\footnotetext[4]{All other cominations have both 
color and hypercolor different from zero. These combinations contain preons and antipreons of
the same type, e.g., $(T{\bar T} V)$, which concerning its quantum numbers are identical to a
single preon.}

They are shown in Table~\ref{tab:2}.
Our assumptions are less restrictive than those of Harari and Seiberg~\cite{har81}. 
These authors suggest that in a ``natural'' theory, nearly massless fermionic bound states 
must be hypercolor singlets. 
\begin{table}[h]
\begin{center}
\begin{tabular}{ l c c c c c c c }
\hline
fermionic group & preon content & bound state &  ${\cal{P}}$ & $\Upsilon$ &  $Q$ & $\Pi$   \\ 
\hline
& & & & & &  \\
 &$\left( VVV \right)$ & $\left( \nu_{e}, \nu_{\mu}, \nu_{\tau} \right)$ & +1 & -1 & 0 & +1 \\
leptons & & & & & & \\
 &$\left(\bar{T}\bar{T}\bar{T} \right)$ & $\left( e^{-}, \mu^{-}, \tau^{-} \right)$ & -1 & -1 & -1 & +1 \\
& & & & & &  \\
\hline
& & & & & & \\
 &$\left( TTV \right)$ & $\left( u, c, t \right)$  & +1 & $+\frac{1}{3}$ & $+\frac{2}{3}$ & +1   \\
quarks & & & & & \\
 &$\left(\bar{T}\bar{V}\bar{V} \right)$ & $\left( d, s, b \right)$ &  -1 & $+\frac{1}{3}$ & $-\frac{1}{3}$
& +1\\ 
& & & & & &  \\
\hline
& & & & & &  \\
 &$\left( TT\bar{V} \right)$ & $\left( \tilde{u},\tilde{c},\tilde{t} \right)$ & $+\frac{1}{3}$ & +1 & 
$+\frac{2}{3}$ & -1 \\  
hyperquarks & & & & & & \\
 &$\left( \bar{T}VV \right)$ & $\left( \tilde{d}, \tilde{s}, \tilde{b} \right)$ & $+\frac{1}{3}$ & -1 
& $-\frac{1}{3}$ & +1\\
& & & & & & \\
\hline
\end{tabular}  
\caption{Allowed three-preon bound states representing leptons, quarks and
hyperquarks and their quantum numbers (see also \cite{har81}). Formally, the
hyperquarks are obtained from the corresponding quarks by interchanging: $ V
\leftrightarrow \bar{V}$.}
\label{tab:2}
\end{center}
\end{table} 

It is evident from Table~\ref{tab:2} that integer values of ${\cal P}$ correspond to
hypercolor singlets (leptons and quarks) whereas integer values of $\Upsilon$ correspond 
to color singlets (leptons and hyperquarks). 
Fractional values of the quantum numbers ${\cal P}$ and $\Upsilon$ correspond to 
open hypercolor or open color charges. Only colored or hypercolored objects can have fractional 
electric charges. 

Furthermore, one can see from the preon content of the quarks and leptons 
that particles and antiparticles enter in a symmetrical way. 
This leads to a different interpretation of the apparent matter-antimatter asymmetry in the universe.
In terms of the standard model with elementary leptons and quarks there is an asymmetry between 
matter and antimatter. On the other hand, at the preon level there is complete symmetry 
between matter and antimatter. For example, atomic hydrogen, which makes up 90$\%$ of the 
visible universe, contains an equal number of preons and antipreons. We hasten to add 
that this does not explain the observed asymmetry on the bound state level, the origin of which
must be looked for in CP-violating interactions.

\section{'t Hooft anomaly constraint and generation number}
\label{sec:4}


A necessary condition for chiral symmetry conservation needed to explain the near masslessness
of fermionic bound states, is the matching of anomalies at the elementary and 
composite levels, known as the 't Hooft anomaly matching condition~\cite{gth79}. 
The anomaly condition involves the difference $D_i^{abc}$ 
\begin{equation}
\label{Dabc}
D_{i}^{abc} = \sum_{i} d_{i(L)}^{abc} - \sum_{i} d_{i(R)}^{abc}   
\end{equation}
of structure constants $d_i^{abc}$ of the lefthanded (L) and righthanded (R) sector of the fermions ($f_{i}$) 
with
\begin{equation}
d_{i}^{abc} = Tr \left(\lambda_{a}\left\{\lambda_{b},\lambda_{c}\right\}\right)_{i},
\end{equation}
where $\lambda_{a}$ are the generators of the corresponding symmetry.

The 't Hooft anomaly matching condition is usually formulated as~\cite{swi82}:
\begin{equation}
\label{bound states and preons}
\sum_{i} D_{i}^{abc}(\mbox{bound states}) = \sum_{j} D_{j}^{abc}(\mbox{preons}).
\end{equation}
Here, the summation on the right hand side extends over the two types of preons and
on the left hand side over the three types of bound state doublets: leptons,
quarks, and hyperquarks (see Table~\ref{tab:2}). 

The following considerations are separately valid for each chiral sector
and lead to a special case of the 't Hooft anomaly condition, namely the equality
of the sum of bound state and preon charges. 
As can be seen from Table~\ref{tab:1} and Table~\ref{tab:2}, 
adding the preon number of the bound states within each doublet leads to a
cancellation for both the lepton and quark doublets, whereas 
the total preon number of the hyperquark doublet and of the fundamental preon doublet are equal.

Similarly, we obtain for the electric charge $Q$ from Tables~\ref{tab:1} and \ref{tab:2}  
\begin{equation}
\label{chargeA1} 
[ Q(\tilde{u}) + Q(\tilde{d}) ] N_{H} N_{G} =  [ Q(T) + Q(V) ] N_{H} N_{C}. 
\end{equation}
From Eq.(\ref{chargeA1}) we conclude that the hyperquarks must occur with a multiplicity of 
$N_G$ (generation number) in order to balance the number of colors on the preon side, i.e., $N_C=N_G$.
For the electric charge $Q$ of the leptons and quarks we get 
\begin{equation}
\label{sm anomaly} 
Q(\nu) + Q(e^{-}) +  [ Q(u) + Q(d) ] N_{C} = 0, 
\end{equation}
which is the usual anomaly freedom condition of the electroweak theory.
It is thus explained by the preon content of these fermions.
Adding Eq.(\ref{chargeA1}) and Eq.(\ref{sm anomaly}) we obtain
\begin{eqnarray}
\label{chargematching}
& & \hspace{-0.9 cm} \Bigl ( Q(\nu) + Q(e^{-})  +  [ Q(u) + Q(d)] N_{C} + \nonumber \\  
& & \hspace{-0.9 cm} [Q(\tilde{u}) + Q(\tilde{d})] N_{H} \Bigr )  N_{G} 
 =  [Q(T) + Q(V)] N_{H} N_{C}.
\end{eqnarray}
This can be written in compact notation as
\begin{equation}
\sum_{i} Q_{i}(\mbox{bound states}) = \sum_{j} Q_{j}(\mbox{preons}).
\end{equation}
Note that this equality only holds if the multiplicity of 
the color/hypercolor degrees of freedom on the preon side is balanced by the generation number 
$N_G$ on the bound state side.

The preceding argument can be repeated by replacing the hyperquarks by quarks in Eq.(\ref{chargeA1}) 
and quarks by hyperquarks in Eq.(\ref{sm anomaly}). We then obtain $N_G=N_H$ from which follows
\begin{equation}
N_{G} = N_{C} = N_{H}.
\end{equation}

\section{Summary and outlook}
\label{sec:5}

A composite model approach of quarks and leptons gives answers to some fundamental
questions, which remain unanswered within the standard model. 
For example, the exact equality of the electric charge of proton and positron can now be 
explained by the preon and antipreon content of these particles. Furthermore, the asymmetry 
between matter and antimatter is now seen as a property of the bound states and is no longer 
related to an asymmetry in the realm of fundamental particles where this symmetry is restored. 
Moreover, the connection between fractional electric charge and 
colored or hypercolored fermionic bound states is to some extent
explained in the preon model.

We have suggested that the 't Hooft anomaly cancellation constraint, which is a necessary 
condition to explain the near masslessness (compared to the preon scale) of quarks and leptons 
can be fulfilled if a new class of preon bound states, called hyperquarks, is introduced. 
We have also suggested that the anomaly cancellation condition then leads to a connection between 
the number of color and hypercolor degrees of freedom and the generation number. This 
restricts the number of fermionic bound states to exactly three generations. 

At what energy scale can one expect the hyperquarks to appear?
Because hyperquarks have not yet been observed with present accelerators,
a lower limit for the hyperquark mass is  $m_{hq} > 1$ TeV. 
On the other hand, above $10^{16}$ GeV, i.e., 
the energy scale of proton decay, quarks transform into leptons due to preon exchange
processes, i.e., explicit preon degrees of freedom become important. 
Because the 't Hooft anomaly condition requires that the preon bound states 
be massless compared to the preon scale, the hyperquark masses 
are presumably closer to the lower experimental limit of 1 TeV.

We hope to discuss bosonic preon bound states in a future publication.

\noindent
{\bf Acknowledgements}
We would like to thank Haim Harari and Don Lichtenberg for helpful
correspondence.


\begin{thebibliography}{9}
\bibitem{comment1}
The hyperquark states are refered to as hyperlepton 
states in Ref.~\cite{har80}. 
Because of their fractional charge these states are more 
appropriately called hyperquarks. The states which Harari and Seiberg~\cite{har80} call 
hyperquark states have the same quantum numbers as the fundamental preons, in particular 
they carry both color and hypercolor and thus are not in the class of allowed bound states  
(see sect. \ref{sec:3}).
\end{thebibliography}
\end{document}